\documentstyle[multicol,epsf,aps,pre,tabularx]{revtex}
\def\figspace{\vspace{0.4cm}}

\begin{document}
\draft

\title{On the critical behavior of a lattice prey-predator model}
\vspace {1truecm}
\author{Tibor Antal${}$ and Michel Droz${}$}
\address{ D\'epartement de Physique Th\'eorique, Universit\'e de
          Gen\`eve, CH 1211 Gen\`eve 4, Switzerland.}
\author{Adam  Lipowski${}$}
\address{Department of Physics, A. Mickiewicz University,61-614 Pozna\'n, 
Poland}
\author{G\'eza \'Odor${}$}
\address{Research Institute for Technical Physics and 
Materials Science, \\ H-1525 Budapest, P.O.Box 49, Hungary}    

\maketitle

\begin{abstract}
  The critical properties of a simple prey-predator model are
  revisited.  For some values of the control parameters, the model
  exhibits a line of directed percolation like transitions to a single
  absorbing state. For other values of the control parameters one
  finds a second line of continuous transitions toward infinite number
  of absorbing states, and the corresponding steady-state exponents
  are mean-field like. The critical behavior of the special point $T$
  (bicritical point), where the two transition lines meet, belongs to
  a different universality class. The use of dynamical Monte-Carlo
  method shows that a particular strategy for preparing the initial
  state should be devised to correctly describe the physics of the
  system near the second transition line.  Relationships with a forest
  fire model with immunization are also discussed.
\end{abstract}

\date{\today}
\begin{multicols}{2}
\narrowtext

\section{Introduction}
\label{sec:intro}
The study of prey-predator systems  has attracted a lot of attention
since the pioneering works of Lotka \cite{lotka}  and Volterra
\cite{volterra}. Working at a mean-field level (homogeneous
populations) they showed that, depending on the initial state, the
system can evolve toward a simple steady-state or a limit cycle, in
which the populations oscillate periodically in time.

An important question is the understanding of the role played by the
local environment on the dynamics (spatial effects) \cite{may} and,
accordingly, many extended prey-predator models have been studied
during the past years
\cite{bascompte,tome,boccara,provata,tainaka,lipowski}.
Recently, a simple prey-predator model was introduced by some of us
\cite{drozal}.  Although governed by only two control parameters, this
model exhibits a rich phase diagram.  As a function of the two control
parameters $\lambda_a$ and $\lambda_b$, which are the growth rates of
prey and predator respectively, two different phases are observed: a
pure prey phase (P), and a coexistence phase of prey and predator in
which an oscillatory (O) region and a non-oscillatory (NO) region can
be distinguished. For a system size $L \to \infty$, these three
different domains meet at a particular point, called $T=(\lambda_a^T,
\lambda_b^T)$ (precise definitions are given below).  It was shown
\cite{drozal} that $\lambda_a^T =0$ and $\lambda_b^T \approx 5.0 \pm
0.3$.  For $\lambda_a > 0$, a phase transition line between the pure
prey phase and the coexistence phase is present, and the critical
exponents along this line are the ones of directed percolation (DP)
\cite{DP}.  However, it was also observed that when the growth rate of
prey $\lambda_a \to 0$ and $\lambda_b > \lambda_b^T$, the model
undergoes a non-DP continuous phase transition. Since the directed
percolation is a generic universality class for models with absorbing
states (unless some special conditions are satisfied~\cite{haye}),
existence of such a transition is certainly surprising. These two
lines of different continuous non-equilibrium phase transition meet at
the bicritical point $T$ \cite{fisher}, and one forecast that the
critical behavior at this particular point may also belong to a new
universality class.

The goal of this paper is to study in more details the properties of
these non-DP phase transitions.  First, we perform extensive
steady-state simulations, which confirm the non-DP character of the
transition in the limit $\lambda_a\rightarrow 0$ and
$\lambda_b\ge\lambda_b^T$.

As it was already shown~\cite{drozal}, in this limit the model
exhibits oscillatory behavior.  But, in addition to that, for
$\lambda_a=0$ the model has infinitely many absorbing states. These
two properties are responsible for a rather peculiar behavior of the
model, which becomes particularly transparent when the model is
examined using the dynamical Monte Carlo method.  When applied to models
with infinitely many absorbing states, this dynamical method uses the
so-called natural absorbing states, which are the most likely states
to be reached by the dynamical evolution of the system.  We show,
however, that this common but somehow heuristic procedure fails here.
Indeed, for the present model, natural absorbing states contain only
short-ranged islands of prey on which spreading is not critical.  To
restore criticality of spreading we generated the absorbing states
using a quasi-static approach to the critical point.  This example
shows that for some models with infinitely many absorbing states a
special approach is needed to examine the dynamical properties of the
critical point.

The paper is organized as follows. In Sec.~\ref{sec:model}, the
model is defined and some of its properties discussed.  A thorough
investigation of the critical behavior has been made using two
different complementary approaches.  In Sec.~\ref{sec:steady}, the
critical behavior is investigated using steady state properties while
in Sec~\ref{sec:dynamic} one uses dynamical Monte Carlo method.
It is shown that for $\lambda_a\rightarrow 0$ and
$\lambda_b\ge\lambda_b^T$ the steady-state exponents are indeed
mean-field-like, while the dynamical exponents are non-universal depending
continously upon $\lambda_b$. Nevertheless, trace of the mean-field
character of the transition shows up in scaling relation among
dynamical exponents. The critical behavior at point $T$ is also
investigated, and it turns out that the corresponding exponents belong
to a new universality class.  Finally, physical arguments explaining
the above findings are given in Sec.~\ref{sec:conc}.

\section{Model}
\label{sec:model}

The model used in Ref. \cite{drozal} is defined as follows.
Each cell of a two-dimensional square lattice (of size $L \times L$,
with periodic boundary condition), labeled by the index $i$, can be at
time $t$, in one of the three following states: $\sigma_i = 0$, 1, and 2.
A cell in state 0, 1, or 2 corresponds, respectively, to a cell which is
empty, occupied by prey, or simultaneously occupied by prey and
predators. The transition rates for site $i$ are
(i) $0 \to 1$ at rate $\lambda_a (n_{i,1}+n_{i,2})/4$,
(ii)$1 \to 2$ at rate $\lambda_b (n_{i,2})/4$, and
(iii) $2 \to 0$ at rate 1,
where $n_{i,\sigma}$ denotes the number of nearest neighbor sites of
$i$ which are in the state $\sigma$. 
The first two processes model the spreading of prey and predators.
The third process represents the local depopulation of a cell due to
overly greedy predators.  The rate of the third process is chosen to
be 1, which sets the time scale, hence $t$, as well as $\lambda_a$ and
$\lambda_b$ are dimensionless quantities.

\begin{figure}[]
\centerline{
        \epsfxsize=8cm
        \epsfbox{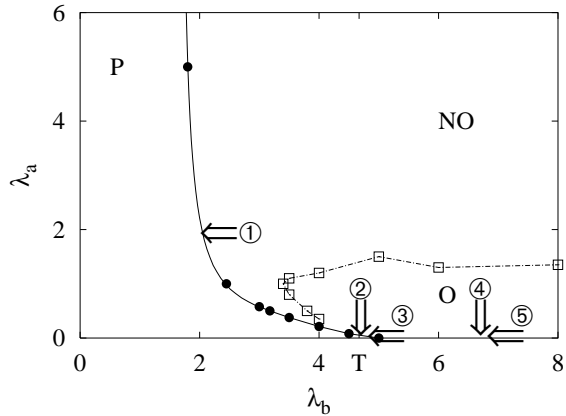}
           }
\figspace
\caption{Phase diagram obtained by extrapolation of the simulation results to
  the case $L\to\infty$. The solid line represents the DP transition
  between the prey phase (P) and the non-oscillatory part (NO) of the
  coexistence phase ( The $\bullet$ symbols are the simulated values
  while the solid line is just a guide to the eyes).  The $\Box$
  symbols delimit the crossover between the oscillatory (O) and
  non-oscillatory (NO) regimes present in the coexistence phase. The
  arrows correspond to the path described in the text, along which the
  critical exponents have been measured.  }
\label{fig:phase_diag}
\end{figure}

The properties of this model have been investigated both by mean-field
and Monte-Carlo methods \cite{drozal}. The Monte-Carlo result
extrapolated to the case $L \to \infty$ is summarized in 
Fig.~\ref{fig:phase_diag}. The transition line between the prey phase and
the coexistence phase, $\lambda_b^*(\lambda_a)$, belongs to the directed
percolation universality class \cite{DP}, as expected, and terminates
at the point $T\equiv(\lambda_a^T=0,\lambda_b^T)$, where the P, O, and
NO domains meet. For $\lambda_b > \lambda_b^T$, the transition between
the oscillatory domain of the coexistence phase and the prey phase 
takes place at $\lambda_a=0$.
Along this transition line the predator density approaches zero as a
power law $b \sim \lambda_a^{\beta_2}$, with $\beta_2 \approx 1$, and
so, does not belong to the DP class, which is somehow unexpected. The
value $\beta_2 \approx 1$ lead to the conjecture \cite{drozal} that
this second transition could be mean-field-like. There is a crossover
between the O and NO parts of the coexistence phase.
The purpose of the present study is to give a complete description of the
nature of the transitions near the line $\lambda_a=0$ for $\lambda_b
\ge\lambda_b^T$.

It is worth to mention, that our model is closely related to a model
introduced by Drossel {\it et al.\ } \cite{drossel} to investigate the
effect of immunization in an extension of the simple forest-fire model
\cite{bak}.  This three-state model (0: empty site, 1: tree, and 2:
burning tree) differs from our model in some details: the growth rate
of a tree ($\sigma : 0\to1$) is $p$, independently of the environment,
and a tree is ignited ($\sigma : 1\to2$) with rate $(1-g)\Theta(n_2)$,
($\Theta$ is the usual Heaviside-function). This second process models
the immunization of trees against fire. The third process ($\sigma :
2\to0$) happens at rate 1. For non zero immunity and $p > 0$ Albano
\cite{albano} showed that the transition toward a single absorbing
state is DP like, while for $p=0$ (at the end point of the DP
transition line), the transition belongs to the dynamical percolation
universality class, and the absorbing state is not unique.

\section{Steady state study of the critical behavior}
\label{sec:steady}

Extensive Monte-Carlo simulations for system sizes up to $4000 \times
4000$ have been performed to investigate the behavior of the predator
density $b$, for $\lambda_a \to 0$ and three different values of
$\lambda_b$, namely, $\lambda_b=4.67$, 5.0, and 6.0, following
trajectories of type 2 and 4 in Fig.~\ref{fig:phase_diag}. The value
$\lambda_a=0, \lambda_b=4.67$ corresponds to the best determination of
the end point T, obtained by the dynamical approach described below.
Owing to the oscillatory behavior near the critical line
($\lambda_a=0,\lambda_b>\lambda_b^T$), the system very easily evolves
into an absorbing state, where the predators are extinct, therefore
careful initialization is needed in the simulations.  Usually $10^4$
MCS "thermalization" were applied following $10^4$ MCS "approaching"
time where $\lambda_a$ was decreased continuously.  The densities
and the fluctuations of prey and predators were averaged over
$\sim 2\times 10^5$ MCS for each $\lambda_a$, $\lambda_b$ points. It
is found that $b \sim \lambda_a^{\beta_2}$ for $\lambda_a \to 0$.  In
order to see corrections to scaling, we compute the effective exponent
\begin{equation}
\beta_{eff}(\lambda_a(i)) = \frac {\ln b(\lambda_a(i)) -
\ln b(\lambda_a(i-1))} {\ln \lambda_a(i) - \ln \lambda_a(i-1)} .
\end{equation}
where $\lambda_a(i)$ and $\lambda_a(i-1)$ are two consecutive values
of the control parameter $\lambda_a$.

\begin{figure}[]
  \centerline{ \epsfxsize=8cm \epsfbox{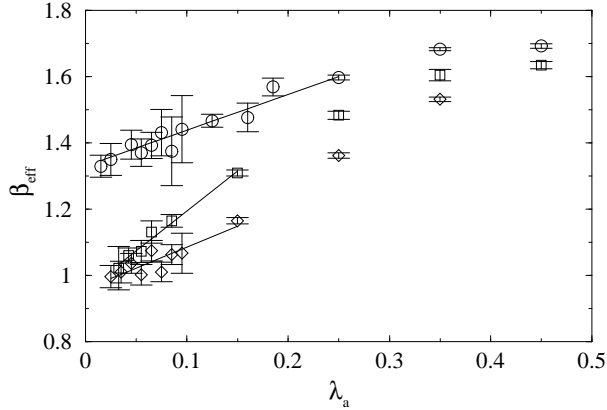} } 
  \figspace
  \caption{Predator density critical exponent $\beta_2$ obtained for 
    several values of $\lambda_a$ and $\lambda_b=4.67$ ($\circ$), 5.0
    ($\Box$) and 6.0 ($\Diamond$).  }
\label{fig:expo_beta}
\end{figure}

As the Fig.~\ref{fig:expo_beta} shows, for $\lambda_b=5$ and 6, the linear
extrapolation of $\beta_{eff}$ converges to $\approx 1$ within
statistical errors
\begin{eqnarray}
\beta_2(\lambda_b=5.0)=1.01(1),  \\
\beta_2(\lambda_b=6.0)=0.96(2).
\end{eqnarray}
For $\lambda_b=4.67$ (trajectory of type 2 in Fig.~\ref{fig:phase_diag}) it goes to a
somewhat higher value, and one finds 
\begin{equation}
  \beta_2(\lambda_b=4.67)=1.33(4).
\end{equation}
The measurement of the fluctuations
\begin{eqnarray}
  \chi_b = L^2 \langle( b - \langle b \rangle)^2 \rangle \ 
  \sim \lambda_a^{-\gamma}
\end{eqnarray}
are less precise and we estimate $\gamma = -0.6(16)$ at $\lambda_b=4.67$,
and $\gamma \sim 0$ for $\lambda_b=5$, 6.

For $\lambda_b > \lambda_b^T$ these values are consistent with the
previous prediction \cite{drozal} $\beta_2 \approx 1$.  However, at
the bicritical point $T$ the value of $\beta_2$ is completely different and
thus belongs to a new universality class.

\section{Dynamical study of the critical behavior}
\label{sec:dynamic}

A very useful technique to study the critical properties of a system
with absorbing states is the so-called dynamical Monte Carlo method
\cite{grass}.  In this approach, the system is prepared in an initial
state, which is one of the absorbing states up to one site, which is set
to be in the active state.  One considers an ensemble of trials
starting from the same initial state. Certain dynamical quantities
exhibit a power law behavior when the system is critical.  For
example, the survival probability behaves as 
\begin{eqnarray} P(t) \sim t^{-\delta}.
  \label{expdelta}
\end{eqnarray}
The deviation from this power law behavior, when the system is
off-critical, provides a very precise way to locate the critical
point.

The number of active sites $N(t)$ behaves as
\begin{eqnarray}
  N(t) \sim t^{\eta}
  \label{expeta}
\end{eqnarray}
while, for the mean square spreading from the origin $R^2(t)$ 

\begin{eqnarray}
  R^2(t) \sim t^z ,
\label{expz}
\end{eqnarray} 
where the dynamical exponent $z=2\nu_{\perp}/\nu_{\parallel}$ is the
ratio of the critical exponents of spatial ($\nu_{\perp}$) and
temporal ($\nu_{\parallel}$) correlation lengths.  Some scaling
relations between these exponents can be also
derived~\cite{dyna_perc}.

We made various simulations using the dynamical Monte Carlo method and our
results are summarized below.

\subsection{Case of $\lambda_a=0$}

\begin{figure}[]
  \centerline{ \epsfxsize=8cm \epsfbox{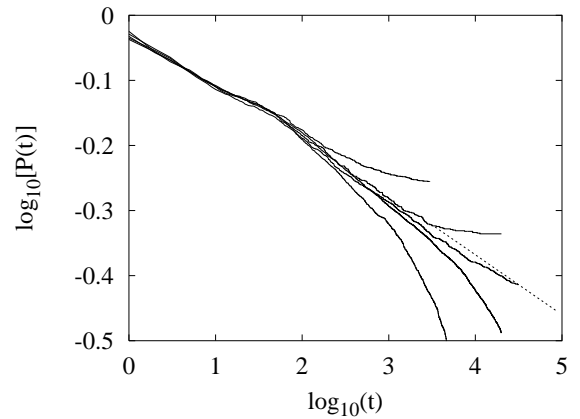} } 
  \figspace
  \caption{The survival probability $P(t)$ as a function of $t$ obtained for
    $\lambda_a=0$ and (from the top) $\lambda_b=4.6$, 4.65, 4.67, 4.7,
    and 4.75 (trajectory of type 3 in Fig.~\ref{fig:phase_diag}).  We
    used system size $L=3000$, and up to $10^5$ independent runs were
    made for each value of $\lambda_b$.  The dotted line has the slope
    corresponding to $\delta=0.092$.  }
\label{dyna0}
\end{figure}


First, we simulated the model on the $\lambda_a=0$ line, taking, as an
absorbing state, a lattice filled with prey (trajectory of type 3 in
Fig.~\ref{fig:phase_diag}).  Measuring the survival probability
$P(t)$, we found that $\lambda_b=\lambda_b^T\approx 4.67$ is the critical
point, which separates the absorbing phase ($\lambda_b<\lambda_b^T$)
and the phase with annular growth ($\lambda_b>\lambda_b^T$). Measuring
the slope at $\lambda_b=\lambda_b^T$ (see Fig.~\ref{dyna0}) we estimate
$\delta \approx 0.095(5)$, which is very close to the value obtained
for dynamical percolation, for which, in two dimensions,
$\delta=0.092$ \cite{dyna_perc}. Moreover, using (7) and (8) we
obtained $\eta=0.60(5)$ and $z=1.72(4)$, which are also very close to the
dynamical percolation values.

Note that the usual $\beta$ exponent for dynamical percolation,
which takes the value $\beta \sim 0.14$, is defined through
$b\sim(\lambda_b^T-\lambda_b)^\beta$, which differs from our
definition of $\beta_2$ in Sec.~\ref{sec:steady}. Thus it is not
surprising that these two exponents differ.  Note also that the
dynamical estimation of the critical endpoint $\lambda_b=\lambda_b^T$
has been used in the static approach of Sec.~\ref{sec:steady}.

\subsection{Case of $\lambda_a>0$}

\begin{figure}[]
\centerline{
        \epsfxsize=8cm
        \epsfbox{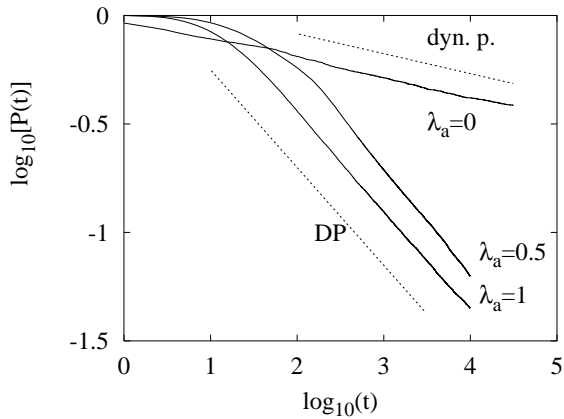}
           }
\figspace
\caption{The survival probability $P(t)$ as a function of $t$ obtained for
  $\lambda_a=0$, $\lambda_b=4.67$ (trajectory of type 3 in
  Fig.~\ref{fig:phase_diag}), for $\lambda_a=0.5$, $\lambda_b=3.175$,
  and for $\lambda_a=1$, $\lambda_b=2.451$ (trajectory of type 1 in
  Fig.~\ref{fig:phase_diag}).  Corresponding values of $\lambda_a$ are
  also shown in the figure.  For $\lambda_a>0$ we used the system size
  $L=1000$, and up to $10^5$ independent runs were made for each
  curve.  The dotted lines have the slope corresponding to the
  exponent $\delta$ of the dynamical and the directed percolation.}
\label{dync}
\end{figure}

The same scheme was used for $\lambda_a>0$.  The critical point
was located for $\lambda_a=0.5$ and 1 (trajectory of type 1 in
Fig.~\ref{fig:phase_diag}), and they agree with the steady state results
of Ref.~\cite{drozal}.  Measuring the slope at criticality (see
Fig.~\ref{dync}) we estimate $\delta \approx 0.45$, i.e., the value
compatible with DP \cite{dyna_perc}. The fact, that for $\lambda_a>0$
the phase transition belongs to the DP universality class, was already
confirmed using the static calculations \cite{drozal}.

In Sec.~\ref{sec:model} we already noticed the similarity of the
present model with a forest-fire model with immunization.  Results
presented in this section provide farther arguments supporting such an
analogy.  Indeed, dynamical exponents measured by Albano for the
forest fire model are also close to the dynamical percolation (without
growth of tree) and the directed percolation (with growth of tree)
\cite{albano}.

\subsection{Inhomogeneous absorbing states}

Static simulations suggest that the model becomes critical on the line
$\lambda_b>\lambda_b^T$ and $\lambda_a=0$.  Moreover, let us notice
that for $\lambda_a=0$ there are infinitely many absorbing states:
indeed, any configuration without predators is an absorbing state.  It
is well known that dynamical Monte Carlo method can be applied also to
models with infinitely many absorbing states.  However, as we will see
below, applicability of this method to the criticality on this line
requires serious reconsiderations.

First, let us recall that the dynamical Monte Carlo method for models
with infinitely many absorbing states usually uses the so-called
natural absorbing states, i.e. states which are reached by the
model's dynamics.  Numerical evidence suggests that for such states the 
dynamical critical point coincides with the static one.  Moreover,
dynamical exponents $\delta$ and $\eta$, measured on such states, take
universal values.

\begin{figure}[]
\centerline{
        \epsfxsize=8cm
        \epsfbox{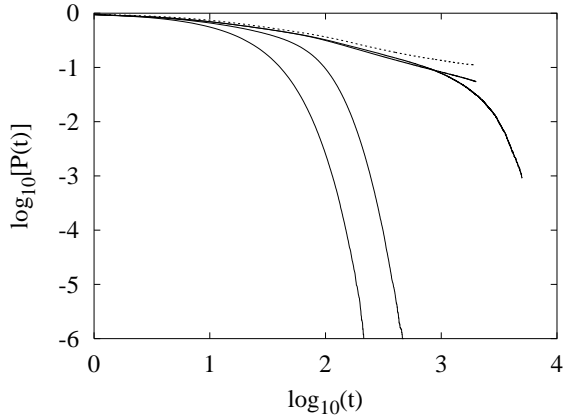}
           }
\figspace
\caption{The survival probability $P(t)$ as a function of $t$ obtained for
  $\lambda_a=0$ and for $\lambda_b=6$ (solid line) and 8 (dotted
  line).  Absorbing states were obtained using continuous cooling with
  the cooling rates (from the top) $r=0.0001$, 0.0001, 0.001, 0.01, and
  $\infty$.  For the slowest cooling we used the system size
  $L=1000$.  For each cooling rate we generated $10^3$ absorbing
  states and for each absorbing state we generated from $10^2$ to $10^5$
  independent runs. }
\label{cool}
\end{figure}

Following this prescription, we generated natural absorbing states for
$\lambda_a=0$ and $\lambda_b>\lambda_b^T$, and then used such states
to perform dynamical simulations.  An initial configuration was chosen
randomly, with equal probabilities of a site being empty, occupied by
prey, or simultaneously by prey and predator.  Fixing $\lambda_a(=0)$
and $\lambda_b$, we then allowed the system to evolve until an absorbing
configuration was reached (i.e. all predators die out).  Our results,
presented in Fig.~\ref{cool}, show, however, that the spreading of
activity is not critical (i.e. power-law), but rather exponential.
But we can argue that this is not surprising.  Indeed, a random
initial configuration (with probability of prey equal 1/3) is below
the percolation threshold with respect to the clusters of prey, and
contains only finite clusters of them~\cite{comment1}.  On such
clusters activity has certainly finite duration (for $\lambda_a=0$
prey do not grow) and the exponential decay of $P(t)$, as it can be
seen in Fig.~\ref{cool}, is an expected feature.

Clearly, the lack of criticality in $P(t)$ is due to the finiteness of
prey clusters in the natural absorbing states.  In principle, we
can cure this effect starting from random initial configurations but
containing larger fraction of prey.  For sufficiently large
concentrations the system will be above the percolation threshold and
activity will be able to spread infinitely. Probably, for a certain
concentration of prey we can tune the system to have power-law decay
for $P(t)$.  Such a procedure, however, is somehow artificial, and
criticality of spreading will not be related with the static
criticality of the system.

\begin{figure}[]
  \centerline{ 
    \epsfxsize=8cm 
    \epsfbox{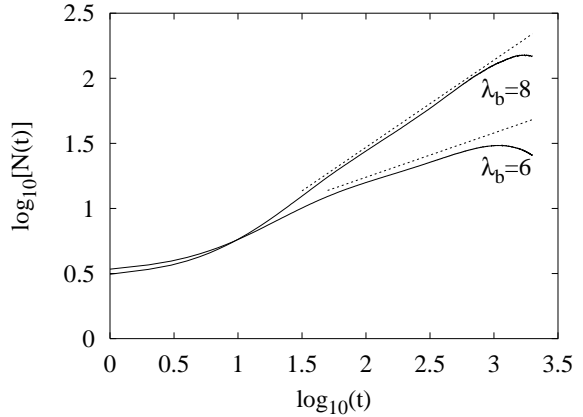} 
    } 
  \figspace
  \caption{The number of active sites $N(t)$ as a function of $t$ obtained for
    $\lambda_a=0$, $\lambda_b=6$, and 8.  Absorbing
    states were obtained using continuous cooling with the cooling
    rate $r=0.0001$.  Straight dotted lines have slopes corresponding
    to $\eta=0.67$ and $\eta=0.35$.  }
  \label{cool1}
\end{figure}

\begin{figure}[]
\centerline{
        \epsfxsize=8cm
        \epsfbox{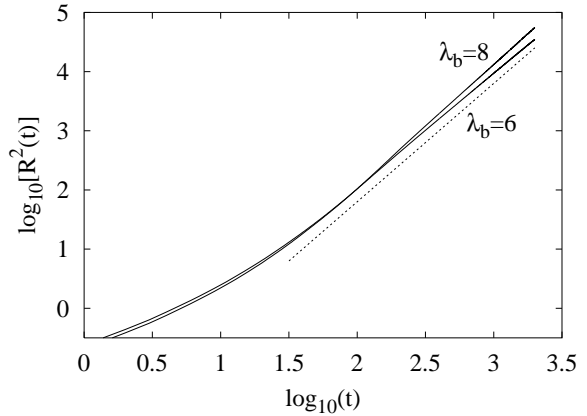}
           }
\figspace
\caption{The squared distanced of active sites $R^2(t)$ as a function of $t$
  obtained for $\lambda_a=0$, $ \lambda_b=6$ and 8.
  Absorbing states were obtained using continuous cooling with the
  cooling rate $r=0.0001$.  The straight dotted line has slope
  corresponding to $z=2$.  }
\label{cool2}
\end{figure}

The question arises here, whether it is possible to generate absorbing
states on which spreading would be critical, and where this criticality
would be generated ``more naturally''?  Hopefully, criticality of
spreading on absorbing states obtained during such a procedure should
be related with the steady-state criticality of the model.  In the
following we suggest a procedure which imitate the quasi-static
approach to the critical point on the line $\lambda_a=0$.  In our
approach we gradually reduce $\lambda_a$ according to the formula:
\begin{equation}
\lambda_a=\lambda_a^0{\rm exp}(-rt),
\label{cooling}
\end{equation}
where $\lambda_a^0=1$ and $r$ is the 'cooling' rate.  [We expect that
the detailed time dependence in Eq.~(\ref{cooling}) is not relevant as
long as it is a slow process].  We terminate the cooling when an
absorbing state is reached.  When the cooling is slow, the system have
enough time to build large clusters of prey.  Our simulations for
$\lambda_b=6$ suggest (see Fig.~\ref{cool}) that in the limit
$r\rightarrow 0$, such absorbing states are critical, with $\delta=
0.59(10)$ (along trajectories of type 5 in Fig.~\ref{fig:phase_diag}).
Measuring the number of active sites $N(t)$, and using
Eq.~(\ref{expeta}) we estimate $\eta= 0.34(10)$ (see
Fig.~\ref{cool1}). The departure of the curves from a straight line
observed for large values of time is related to the finiteness of the
cooling rate.  Moreover, we measured the averaged squared distance
$R^2(t)$, and using Eq.~(\ref{expz}) we obtain $z=2.0(1)$ (see
Fig.~\ref{cool2}).  Actually, we expect that the correct value of this
exponent is $z=2$.  Indeed, in Eq.~(\ref{expz}) one makes the average
only over surviving runs, thus the long-time contributions to $R^2(t)$
come from the rare events, when the activity happened to be placed on
a large island of prey.  Since $\lambda_b>\lambda_b^T$, the activity
on such islands spreads in a deterministic way (annular growth), which
leads to $z=2$.


The same procedure for $\lambda_b=8$ yields: $\delta=0.35(10)$,
$\eta=0.67(10)$ and $z=2.0(1)$.  (Relatively large errors of
estimations of critical exponents are due to several, and difficult to
estimate factors, as finite time $t$, finite cooling rate $r$, and
statistical fluctuations.)  Such results confirm that $z=2$, and
suggest that exponents $\delta$ and $\eta$ might change continuously
with $\lambda_b$.  Non-universality of these exponents is a well-known
property for some other models belonging to the directed percolation
universality class \cite{jose}.  Note, however, that the situation is
different in our case, because the non-universality is related to the
value of $\lambda_b$, rather than to the choice of the initial state.
Such control parameter dependence has already been observed in other
models \cite{odor}. Note that non-universal behavior is not present
along the DP line since the corresponding absorbing state is unique.
Let us finally note that $\delta+\eta$ seems to be close to unity,
which is an exact mean-field result ($\delta_{MF}=1, \eta_{MF}=0$).
This is the only dynamical trace of the mean-field nature of the
transition observed in the steady state.  Let us emphasize, however,
that the criticality of spreading appears only if we prepare the
absorbing states using the method which mimics the quasi-steady-state
evolution of the model.

\section{Conclusions}
\label{sec:conc}
The detailed investigation of the critical properties of a simple
prey-predator model introduced in Ref.~\cite{drozal} showed the
presence of three different types of non-equilibrium phase transitions
between active and absorbing states.  First, the existence of a usual
DP-like transition line was confirmed at $\lambda_b^*(\lambda_a)$ for
$\lambda_a>0$.  Second, a mean-field-like transition was observed for
$\lambda_a \to 0$, $\lambda_b > \lambda_b^T$. The mean-field character
of this transition can be explained in terms of oscillations present
in the model. Indeed, as described in Ref.~\cite{drozal}, when
$\lambda_a \to 0$, the system is subject to large density
oscillations. These oscillations are generating an important local
mixing of the possible states, leading to a mean-field-like behavior.
The criticality along the $\lambda_a=0$ line was confirmed with the
dynamical approach using specially prepared inhomogeneous initial
states.  Some dynamical trace of the mean-field nature of this
transition was also observed. Third, at the bicritical point $T$,
where the two different critical lines meet, we found a dynamical
percolation type transition moving along the $\lambda_a=0$ line, while
approaching the $T$ point from finite $\lambda_a$'s we observed a new
type of critical behavior. 

The measured exponents corresponding to the
$\lambda_a=0$ line are summarized in Table \ref{tab:exps}.
The best numerical estimates for the critical exponents of the
two-dimensional dynamical percolation are given for comparison:
$\delta=0.092$, $\eta=0.586$, and $z=1.771$ \cite{dyna_perc}.

\acknowledgements This work has been partially supported by the Swiss
National Foundation, the Hungarian Academy of Sciences (Grant OTKA
T-25286, T-029792 and Grant Bolyai BO/00142/99), and by the Polish
State Committee for Scientific Research (5 P03B 032 20).  The
simulations were performed partially on the SZTAKI parallel computing
cluster.


\begin{table}[h]
\begin{tabular}{c c c c c} \hline \hline
  Exponent~~~~    & $\lambda_b=4.67~~$ & 5.0~~   & 6.0~~     & 8.0~~ \\
  \hline
  $\beta_2$         & 1.33(4)            & 1.01(1) & 0.96(4) & -- \\
  $\gamma$        & -0.65(10)          & -0.1(1) &-0.05(5) & -- \\ \hline
  $\delta$        &  0.095(5)          & --      & 0.59(10)& 0.35(10) \\
  $\eta$          &  0.60(5)           & --      & 0.34(10)& 0.67(10) \\
  $z$             &  1.72(4)           & --      & 2.0(1)  & 2.0(1)  \\
  \hline \hline\\
\end{tabular}
\caption{Critical exponents around $\lambda_a=0$}
\label{tab:exps}
\end{table}

\end{multicols}

\begin{references}
\bibitem{comment1} Actually, we expect that clusters of prey of any
  size might be formed during such a procedure but large clusters most
  likely appear with exponentially small (with the cluster size)
  probability.  Clusters of such a small probability of appearance
  provide negligible contributions to $P(t)$.
\end{references}
\end{document}